\documentclass{article}
\usepackage[T1]{fontenc} % add special characters (e.g., umlaute)
\usepackage[utf8]{inputenc} % set utf-8 as default input encoding
\usepackage{ismir,amsmath,cite,url}
\usepackage{graphicx}
\usepackage{color}
\usepackage{CJKutf8}
\usepackage{multirow}  % 多行合并的宏包
\usepackage{algorithm}
\usepackage{algpseudocode}
\usepackage{amsmath}
\title{Playing Technique Detection by Fusing Note Onset Information in Guzheng Performance}

\multauthor
{Dichucheng Li$^1$ \hspace{1cm} Yulun Wu$^1$ \hspace{1cm} Qinyu Li$^2$ \hspace{1cm} Jiahao Zhao$^1$} { \bfseries{ Yi Yu$^3$ \hspace{1cm} Fan Xia$^2$ \hspace{1cm} Wei Li$^{1,4}$}\\
$^1$ School of Computer Science and Technology, Fudan University, Shanghai, China\\
$^2$ College of Experimental Art, Sichuan Conservatory of Music, Sichuan, China\\
$^3$ National Institute of Informatics (NII), Tokyo, Japan\\
$^4$ Shanghai Key Laboratory of Intelligent Information Processing, Fudan University, China\\
{\tt\small 21210240219@m.fudan.edu.cn, \{20110240018, 20210240306, weili-fudan\}@fudan.edu.cn, } \\
{\tt\small yiyu@nii.ac.jp, \{935362804, 409769992\}@qq.com}
}

\def\authorname{D. Li, Y. Wu, Q. Li, J. Zhao, Y. Yu, F. Xia, and W. Li}

\begin{document}
\begin{CJK*}{UTF8}{gbsn}

\maketitle
\begin{abstract}
The Guzheng is a kind of traditional Chinese instruments with diverse playing techniques. Instrument playing techniques (IPT) play an important role in musical performance. However, most of the existing works for IPT detection show low efficiency for variable-length audio and provide no assurance in the generalization as they rely on a single sound bank for training and testing. In this study, we propose an end-to-end Guzheng playing technique detection system using Fully Convolutional Networks that can be applied to variable-length audio. Because each Guzheng playing technique is applied to a note, a dedicated onset detector is trained to divide an audio into several notes and its predictions are fused with frame-wise IPT predictions. During fusion, we add the IPT predictions frame by frame inside each note and get the IPT with the highest probability within each note as the final output of that note. We create a new dataset named GZ\_IsoTech from multiple sound banks and real-world recordings for Guzheng performance analysis. Our approach achieves 87.97\% in frame-level accuracy and 80.76\% in note-level F1-score, outperforming existing works by a large margin, which indicates the effectiveness of our proposed method in IPT detection.
\end{abstract}
\section{Introduction}\label{sec:introduction}

\begin{figure}[htbp]
\centering
\includegraphics[width=8.1cm]{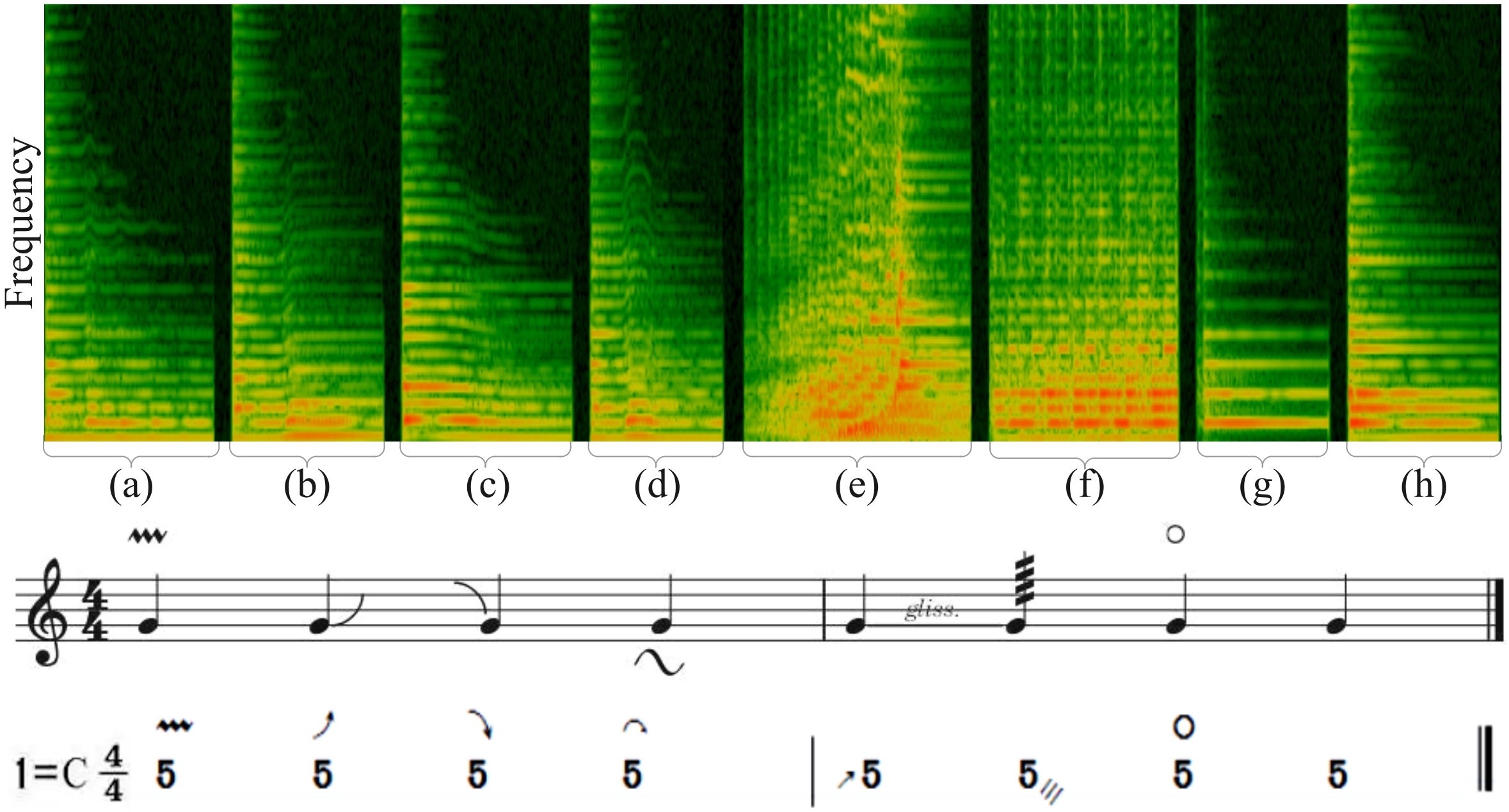}
\caption{The spectrogram, staff notation and numbered musical notation of a Guzheng phrase that contains 8 notes with different playing techniques: vibrato (a), upward portamento (b), downward portamento (c), returning portamento (d), glissando (e), tremolo (f), harmonic (g), plucks (h) }
\end{figure}

The Guzheng (古筝), which is also known as the Chinese zither, is a plucked 21-string Chinese musical instrument existing for over 2,500 years \cite{1}. Chinese traditional music attaches great importance to the melody, so a large number of playing techniques are used to enhance the vividness of Guzheng performance. The pitch variation produced by pressing the strings with the left hand is even regarded as “the soul of Guzheng music” \cite{2}. Instrument playing technique (IPT) detection aims to classify the types of IPTs and locate the associated IPT boundaries in an audio clip. However, there were only few researches about IPTs in the field of Music Information Retrieval (MIR). Particularly, most researches on automatic music transcription (AMT) only consider pitch estimation, while a complete transcription of a Guzheng performance should contain the notations of the playing techniques as shown by the Guzheng numbered musical notation\footnote{a musical notation system widely used in China and called {\itshape Jianpu} in Chinese} in Figure 1. In this work, we propose an IPT detection method that can further be incorporated into a full transcription system of Guzheng music.

One major difficulty for the IPT research is the lack of IPT sound databases. Most of the researches \cite{4,7,8,9} on IPT recognition limited their experiments to samples from isolated notes in a single sound bank. However, according to \cite{13}, the performance of classifiers trained and tested with a single sound bank provides no assurance in the generalization capabilities of the classifiers. Ducher et al. \cite{14} extended the above theory to IPT recognition research. They used five available IPT sound banks in their experiments but did not annotate any recorded corpora of audio.

Existing methods for IPT detection are mainly implemented through two steps\cite{8,9,10}: locating and classifying. These methods are sensitive to the errors caused in locating. Recently, some end-to-end methods\cite{11} have been proposed, while they are weak for variable-length audio and show low accuracy at the boundary of adjacent notes.

Note onset detection aims to localize the very beginning of each note. The beginning of a Guzheng note is easier to identify because the amplitude of that note is at its peak. In Guzheng performance, strings are usually plucked using special nails attached to fingers bottoms. When string is plucked, it has a pre-attack before the string reaches its full level vibration so the onset has a unique broadband spectrum. As shown in Figure 1, each Guzheng playing technique is applied to a note and each note in Guzheng music starts with a clear onset. So onset information is crucial to the Guzheng playing technique detection. However, to the best of our knowledge, no research has taken use of the onset information in the end-to-end IPT detection field.

The main contributions of this paper are as follows: 1) We create a new dataset, GZ\_IsoTech, which consists of Guzheng playing technique clips from two Guzheng sound banks and real-world recordings recorded by a professional Guzheng performer; 2) We propose the first end-to-end method that can be applied to variable-length audio for Guzheng playing technique detection; 3) We propose a decision fusion method where an onset detector is trained to divide an audio into several notes and the predicted IPTs produced by the IPT detector are added frame by frame inside each note to get the IPT class with the highest probability within each note as the final output of that note.

\section{Related Work}

Although the research on IPT detection is still in its early stage, we can summarize relevant researches and divide the development of this field into three periods.

The researches in the first period mainly focus on the playing technique classification of isolated notes\cite{6,7}. However, in reality, there are often continuous notes with varying playing techniques in an audio. We not only need to classify the IPT types but also locate the IPT boundaries. 

The methods for IPT detection proposed in the second period can be divided into two steps: locating and classifying. In \cite{8,9}, signal processing methods were used to select candidates of the playing techniques in the melody contours extracted from 42 electric guitar solo tracks. Then the timbre and pitch features of the candidates were input into the classifiers such as Support Vector Machine (SVM). The candidates were manually selected based on the melody contour, so the methods are sensitive to errors in melody extraction and can hardly be generalized to other IPTs.

In \cite{10}, the candidates of guqin playing techniques were firstly located according to the IPT onset and duration annotations in 39 guqin solo recordings and then classified into six left-hand IPT types. Yet, we generally do not have onset and note duration as prior information in reality. 

In \cite{24}, two binary classifiers bases on Convolutional Neural Networks (CNN) are firstly used to decide whether a fixed-length portion in a piano recording is played with the sustain pedal. Then sliding windows are used to apply the method to pieces with variable lengths. In \cite{14}, five different methods were proposed to classify 18 different IPTs in the cello solo audio concatenated by isolated cello notes with different IPTs from 5 different sound banks. Although the models proposed in \cite{14,24} can detect the IPTs in audio with continuous IPTs, the essence of them is to classify IPTs in a single block. These methods have redundancy in the computation because the input sound is split into overlappping frames that are fed individually to the network through a sliding window. Each convolution is thus computed several times on the overlapping parts.

We regard end-to-end IPT detection as the approach in the third period. In \cite{11}, the authors presented an end-to-end method based on Fully Convolutional Networks (FCN) to detect the IPTs in 10-second segments concatenated by isolated erhu notes. However, the method shows low accuracy at the boundary of adjacent notes and has computational redundancy when applied to variable-length audio.

Our model is built upon the FCN model presented in \cite{11}, with some improvements. We add a dedicated onset detector to take advantage of the significance of note onsets in Guzheng music. A decision fusion is implemented for the onset and IPT prediction to make a note-level prediction. In this way, we achieve a better performance for IPT predictions at the note boundaries. To apply our method to audio with variable lengths, 1D max-pooling layers which only halve the length of frequency axis are used.

\section{GZ\_IsoTech dataset}
In this section, we introduce the Guzheng playing techniques considered in our work and the process of making the dataset.

\subsection{Guzheng playing techniques}
The traditional Guzheng playing techniques can be divided into two classes: plucking string with the right hand and bending string with the left hand \cite{5}. In our work, we consider the following eight playing techniques which are the most frequently used in Guzheng solo compositions.

\textbf{Left-hand playing techniques:}
\begin{itemize}

\item[$\bullet$] \textbf{Vibrato (chanyin颤音):} the periodic oscillation of tones caused by the periodic pressing of a string with left hand (region (a) of Figure 1).

\item[$\bullet$]\textbf{Upward Portamento (shanghuayin上滑音) ({\itshape UP} for short thereinafter):} press a string with left hand to increase the pitch of a ringing note to a desired pitch within major third (region (b) of Figure 1).

\item[$\bullet$]\textbf{Downward Portamento (xiahuayin下滑音) ({\itshape DP} for short thereinafter):} decrease the pitch of a ringing note by releasing a bended string (region (c) of Figure 1). It is the opposite of Upward Portamento.

\item[$\bullet$]\textbf{Returning Portamento (huihuayin回滑音) ({\itshape RP} for short thereinafter):} pluck a string, press it with left hand, and then release it, creating an up-down slide (region (d) of Figure 1).
\end{itemize}

\textbf{Right-hand playing technique: }
\begin{itemize}
\item[$\bullet$]\textbf{Glissando (guazou刮奏, huazhi花指):} play several strings consecutively up or down with thumb or index finger rapidly (region (e) of Figure 1).

\item[$\bullet$]\textbf{Tremolo (yaozhi摇指):} pluck a single string in rapid succession by “shaking” the index finger or thumb of the right hand back and forth across it as shown in the region (f) of Figure 1.

\item[$\bullet$]\textbf{Harmonic (fanyin泛音):} tap a special location of a string with the left hand, and lift the left hand while playing the string with the right hand to produce a corresponding harmonic effect as shown in the region (g) of Figure 1.

\item[$\bullet$]\textbf{Plucks (gou勾, da打, mo抹, tuo托…):} as shown in the region (h) of Figure 1, plucks are defined as simply plucking a string with a finger, including gou, da, mo, tuo\footnote{gou, da, mo, tuo are the names of the Guzheng playing techniques for simply plucking a string using the middle finger, ring finger, index finger, and thumb respectively. }, and so on.
\end{itemize}

\begin{table}
 \begin{center}
 \begin{tabular}{|c|c|c|c|c|}
  \hline
  \multirow{2}{*}{IPT} & \multicolumn{2}{c|}{VSB} & \multicolumn{2}{c|}{RW} \\
  \cline{2-5}
  & num & seconds & num & seconds\\
  \hline
  vibrato & 192 & 261.1 & 42 & 51.5\\
  \hline
  {\itshape UP} & 488 & 752.0 & 48 & 94.4\\
  \hline
  {\itshape DP} & 333 & 508.5 & 51 & 87.5\\
  \hline
  {\itshape RP} & 272 & 327.0 & 94 & 94.7\\
  \hline
  glissando & 316 & 595.3 & 42 & 95.9\\
  \hline
  tremolo & 205 & 259.8 & 23 & 59.7\\
  \hline
  harmonic & 318 & 305.8 & 61 & 42.0\\
  \hline
  plucks & 204 & 204.0 & 135 & 99.7\\
  \hline
 \end{tabular}
\end{center}
 \caption{Quantity of data in the virtual sound banks (VSB) and real-world (RW) recordings. The Guzheng playing technique clips from VSB is used in the training set and that from RW is used in the test set.}
 \label{tab:example}
\end{table}

\subsection{Data collection and labelling}

In order to ensure the diversity of the data, we collected audio clips of single playing techniques (we called these audio clips {\itshape short clips} thereinafter) from real-world recordings (RW) and 2 virtual Guzheng sound banks (VSB) which had different players and recording setups: Yellow River Sound and Kong Audio 3 KONTAKT. We sampled 2328 {\itshape short clips} covering almost all the tones in the range of Guzheng from the sound banks and annotated them into 8 classes one by one according to the sound bank manuals. Their durations range from about 0.6 to 11.5 seconds, in total 3213.4 seconds (see Table 1). The 496 real-world recordings of {\itshape short clips}, in total 625.3 seconds, were recorded and annotated by a professional Guzheng performer. 

We use the {\itshape short clips} sampled in VSB as the training split and the set of {\itshape short clips} recorded in real world as the test split. This division well simulates the test situation in the real world as we usually do not have access to recordings of a testing Guzheng at training time. The dataset with detailed information and demos is available on the website\footnote{https://ccmusic-database.github.io/en/database/ccm.html\#GZTech}\cite{database}. 

\begin{figure*}[htbp]
\centering
\includegraphics[width=16.7cm]{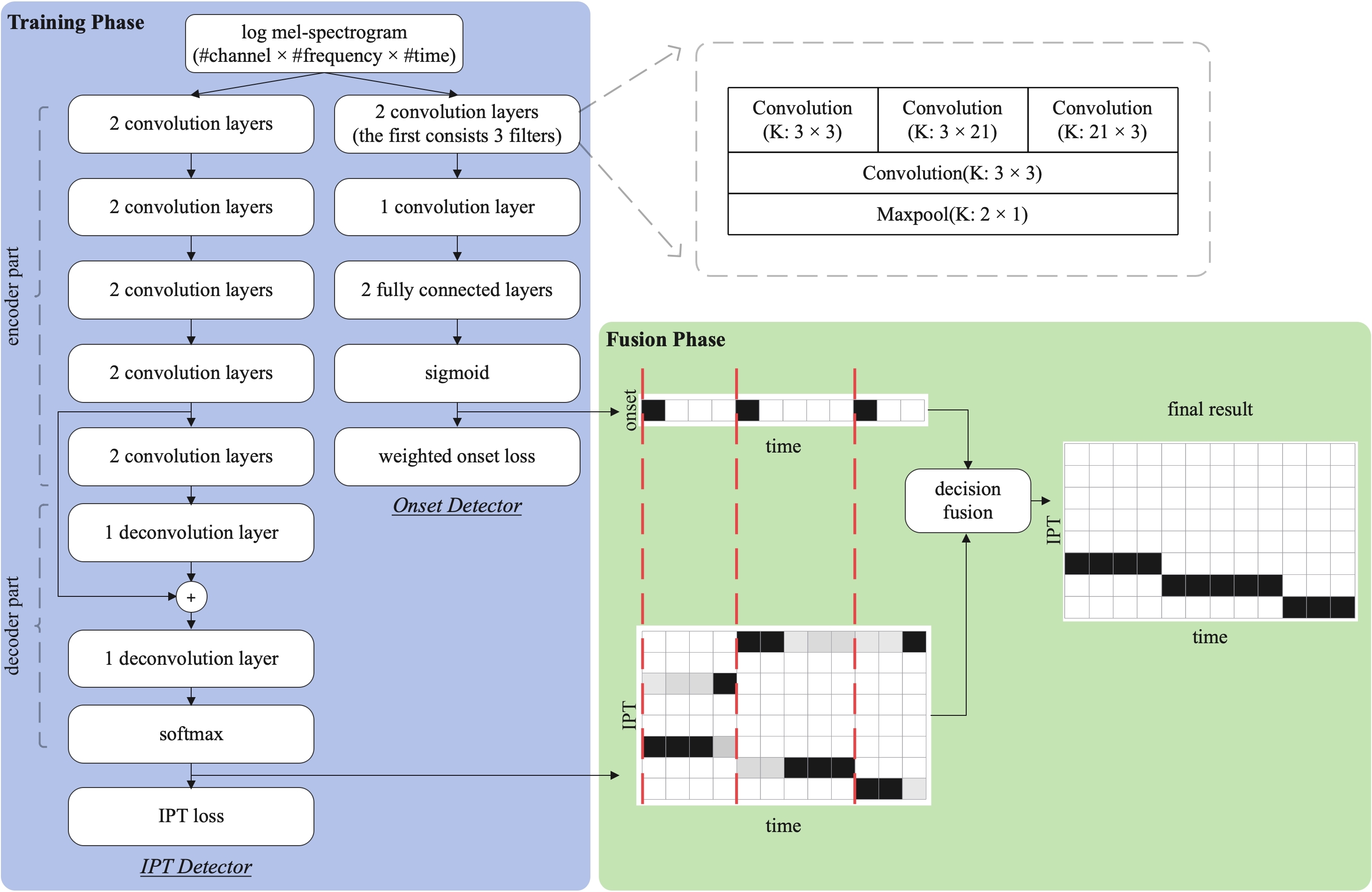}
\caption{System architecture of our proposed model. In the training phase, the IPT detector and the onset detector were trained separately and then decision fusion was applied to their outputs in the fusion phase to get the final result. The vertical red dotted lines in the "fusion phase" stand for the starts of all onset predictions.}
\end{figure*}

\section{METHODS}\label{sec:METHODS}

As shown in Figure 2, our proposed model consists of two modules: the IPT detector and the onset detector. In the training phase, we train the onset detector and the IPT detector separately. Then, in the fusion phase, we apply decision fusion to the thresholded output of the onset detector and the raw output of the IPT detector to get the final result.

\subsection{Input Representation}
Since our goal is to generalize to actual solo recordings, we have generated series of audio sequences which simulate such recordings. Firstly, on both training and test, we concatenated our {\itshape short clips} in the dataset mentioned in Section3 one after another randomly until the length of the concatenated audio sequence is greater than 12.8 seconds.  A check operation was executed to confirm that every generated audio sequence is unique. A 50 milliseconds cross-fade was made in each boundary of adjacent {\itshape short clips} to ensure the audio sequence sound more realistic. We cut the audio sequence generated from the training set directly to 12.8 seconds long for training and use the unsplit audio sequence whose length is greater than 12.8 seconds generated from the test set for testing.

We also label the start time and duration of the corresponding playing techniques. Then we generate two kinds of labels (the onset lables and the IPT labels) by quantizing the continuous time labels into timestamps in the unit of 0.05-second-long frame. 

The input representation is a log mel-spectrogram. Following \cite{11}, we use {\itshape librosa} \cite{28} to compute log mel-spectrogram with 128 logarithmically-spaced frequency bins, an FFT window of 2048, a hop size of 2205, and a sample rate of 44.1kHz. 

\subsection{IPT Detector}\label{subsec:IPT Detector}

Guzheng playing technique detection can be regarded as a part of the task of Guzheng music transcription. As most of the transcription task \cite{16,26}, we treat the frame as a basic unit so we need to identify the playing techniques frame by frame. We can treat the task as a semantic segmentation task since there is no overlap of playing techniques in our data and frames in music are similar to pixels in images. Inspired by the success of FCN in semantic segmentation \cite{18}, which is also extended to audio processing research \cite{11,19}, we applied FCN in our task.

As shown in Figure 2, the IPT detector can be divided into the encoder and the decoder. The encoder part consists of five convolution modules. Two identical convolutional layers were stacked in each module and each of them has square (3 × 3)\footnote{2D filters are noted: N frequency bins x M time frames.} filters. At each convolutional layer, zero padding was applied to make the output have the same length as the input and the output of every convolutional layer was then passed through a Rectified Linear Unit (ReLU). To apply our model to variable-length audio, a 1D max-pooling layer which only halve the length of frequency axis was implementd at the output of each convolution module followed by a dropout layer with the probability 0.25 to aid generalization. The output of the encoder part is the input of the decoder part which is made of two deconvolution layers. The decoder part is used for upsampling the feature map to the shape that we need and then make a frame-level prediction to the IPT. A skip connection of element-wise adding was used from the output of the fourth module in the encoder part to the first module output in the decoder part to help the decoder fuse the information.

\subsection{Onset Detector}
The onset detector is composed of two convolution modules, followed by two fully connected layers. The last layer is a fully connected sigmoid layer with 1 output for representing the probability of an onset for this frame. It has been suggested in recent research \cite{23} that using different musically motivated filter shapes in the first convolutional layer of CNN could improve the model performance. As we discussed in Section 3, different playing techniques in Guzheng vary a lot in the temporal and spectral domain. Thus, we apply three filters of different shape (3 × 3, 3 × 21, 21×3) to the first convolutional layer and the 3 × 21 filter is designed to better capture the feature of long playing technique such as {\itshape portamento} or {\itshape glissando}. The following two convoliutional layers each has 3 × 3 filters.

We use weighted binary cross entropy (WBCE) loss as the loss function for our onset detector.

\begin{footnotesize}
\begin{equation}\label{relativity}
L_{onset} = mean(\sum\limits_{t=0}\limits^{T}{WBCE(\beta, x_t, y_t)})
\end{equation}

\begin{equation}\label{relativity}
WBCE(\beta, x_t, y_t) = \beta y_t\log{x_t} + (2-\beta)(1-y_t)\log(1-x_t)
\end{equation}
\end{footnotesize}where $T$ is the number of frames in the example, $y_t$ is the label that is 1 when there is a ground truth onset at the frame $t$, $x_t$ is the probability output by the model at frame $t$ and $\beta$ is the weight factor of the positive samples. Recent models in the onset detection research showed a lower recall than precision \cite{26,16} and it is largely caused by class imbalance that positive samples for onset are far less than the negative samples. So we decided to increase the weight of the positive samples to balance the performance of the model. We set $\beta$ to $1.94$ by coarse hyper-parameter search.

\begin{table*}
 \begin{center}
 \setlength\tabcolsep{7mm}{
     \begin{tabular}{|c|c|c|c|c|}
      \hline
      \multirow{2}{*}{model} & \multicolumn{1}{c|}{Frame-level} & \multicolumn{3}{c|}{Note-level}  \\
      \cline{2-5}
      & accuracy & precision & recall & F1-score\\
      \hline
      FCN+Onsets & \textbf{87.97} & \textbf{78.20} & \textbf{83.78} & \textbf{80.76}\\
      \hline
      Z.Wang's model\cite{11} & 69.44 & 35.18 & 47.45 & 39.53\\
      \hline
      J.-F. Ducher's model\_Reproduced\cite{14} & 66.93 & 19.37 & 21.31 & 20.05\\
      \hline
      B.Liang's model\_Reproduced\cite{24} & 53.98 & 41.38 & 44.50 & 42.81\\
      \hline
    \end{tabular}
    }
 \end{center}
 \caption{Frame-level accuracy and note-level precision, recall and F1-score on GZ\_IsoTech dataset. Note-based scores calculated by the {\itshape mir\_eval} library. Final metric is the mean of scores calculated per piece in the test set.}
 \label{tab:example}
\end{table*}

\subsection{Fusion of IPT Detector and Onset Detector}

In evaluation, the IPT detector and the onset detector are firstly used separately. Then we use a threshold of 0.5 to make the onset output binary. We separate the IPT output into several clips by the binary onset output. Then we selected an IPT as the final result of each clip by "voting" within it as described by Algorithm 1. We set $n$ as the number of IPTs and $t$ as the number of time frames. $D_{onset}$ whose shape is [t] is the binary output of the onset detector. $D_{IPT}$ whose shape is [$n$, $t$] is the raw output of the IPT detector, representing the probability of each IPT for each frame. $D$ is the final one-hot result that implies the IPT at every frame. There are two intermediate variables in our algorithm. $V$ whose shape is [$n$] represents the total "voting" score for each IPT in the current clip and $b$ denotes the frame number where the current clip starts.

\begin{algorithm}[h]
  \caption{Decision fusion}
  \hspace*{0.02in} {\bf Input:} %算法的输入， \hspace*{0.02in}用来控制位置，同时利用 \\ 进行换行
the number of IPTs $n$, the number of time frames $t$, the thresholded output of the onset detector $D_{onset}$, the output of the IPT detector $D_{IPT}$\\
\hspace*{0.02in} {\bf Output:} %算法的结果输出
final result $D$
  \begin{algorithmic}[1]
    \State{$b \leftarrow 0$}
    \State{$D \leftarrow zeros(D_{IPT})$}
    \State{$V[0,…,n-1] \leftarrow 0$}
    \For{$i\leftarrow0$ \textbf{to} $t-1$}
        \For{$j\leftarrow0$ \textbf{to} $n-1$}
            \State{$V[j] \leftarrow V[j] + D_{IPT}[j][i]$}
        \EndFor
        \If{$i+1 = t$ or $D_{onset}[i+1] = 1$}
            \State{$D[\mathrm{argmax}(V)][b,…,i] \leftarrow 1$}
            \State{$V[0,…,n-1] \leftarrow 0$}
            \State{$b \leftarrow i+1$}
        \EndIf
    \EndFor
    \State \Return $D$
  \end{algorithmic}
\end{algorithm}

\section{EXPERIMENTS}
In this section, we introduce the evaluation metrics we used, the details of the experiment and the result analyses.
\subsection{Evaluation Metrics}
The metrics used to evaluate a model consist of frame-level and note-level metrics. Note-level metrics include precision, recall, and F1 scores. They are calculated by the {\itshape mir\_eval} library \cite{22}. We require that both the IPT type and the onset are correct. In detail, an onset estimation is considered correct if it is within ±0.05 seconds of the ground-truth. Frame-level accuracy is calculated by comparing the output of our model to the ground truth label frame by frame. More specifically, we calculated the ratio of the number of correct frames to the total number of frames as frame accuracy. Both note and frame scores are calculated every audio sequence and we calculated the mean of these scores as the final metric.

\subsection{Results}
To examine whether our proposed architecture (namely \texttt{FCN+Onsets}) can better detect IPTs in audio with continuous IPTs, we reproduced J.-F. Ducher's model \cite{14}, B.Liang's model \cite{24} and Z.Wang's model \cite{11} for comparison, while adapting their capacity to our data.

Because Z.Wang's model \cite{11} cannot be tested on variable-length audio, we cut all the audio from the test set to 12.8 seconds long for the test of Z.Wang's model \cite{11} while other models are still tested on variable-length data.

\begin{table*}
 \begin{center}
 \setlength\tabcolsep{7mm}{
     \begin{tabular}{|c|c|c|c|c|}
      \hline
      \multirow{2}{*}{model} & \multicolumn{1}{c|}{Frame-level} & \multicolumn{3}{c|}{Note-level}  \\
      \cline{2-5}
      & accuracy & precision & recall & F1-score\\
      \hline
      FCN+Onsets & \textbf{87.97} & 78.20 & \textbf{83.78} & \textbf{80.76}\\
      \hline
      CNN+Onsets & 60.80 &45.01 & 48.30 & 46.51\\
      \hline
      No Onset Fusion & 76.48 & 27.74 & 50.55 & 35.47\\
      \hline
      Onset3×3 & 87.70 & 72.67 & 82.76 & 77.19\\
      \hline
      No Skip Connection & 82.45 & 71.05 & 76.15 & 73.39\\
      \hline
      No Weighted Loss & 86.26 & \textbf{80.88} & 79.46 & 80.03\\
      \hline

    \end{tabular}
    }
 \end{center}
 \caption{Ablation studies with frame-level accuracy and note-level precision, recall and F1-score on GZ\_IsoTech dataset.}
 \label{tab:example}
\end{table*}

Table 2 shows the frame-level and note-level results of Guzheng playing technique detection on the GZ\_IsoTech dataset. Our \texttt{FCN+Onsets} model reaches 87.97\% in frame-level accuracy and 80.76\% in note-level F1-score, producing the best scores in both frame-level and note-level metrics. Besides, our proposed model outperforms other models by a large margin in all metrics and results in over a 100\% relative improvement in note-level F1-score compared to other models. This is mainly because our model takes better advantage of the note onset detection results which is highly interrelated to the note-level metrics. Because the test set consists of audio of real-world recordings that is different from training set in performer, instrument and recording environment, the results show that our model offers good generalization capabilities.

The visualization of the input spectrogram, the thresholded onset output, the raw IPT output, the final fusion result and the label for a recording from the test set are shown in Figure 3. Through the images in Figure 3, we can clearly see the importance of fusing IPT activations with onset predictions. The vertical red lines in the third and fourth images are the onset predictions. We can separate the audio into several notes using the onset predictions. Each note in our dataset only corresponds to one playing technique, but there may be several IPT predictions in one note as shown in the "Raw IPT Output" image. So we implement the decision fusion method to choose one IPT as the final output of a note as illustrated in Algorithm 1. The IPT results after being restricted by the onset are presented in the second-to-bottom image ("Final Fusion Result"). The "Target" image is the visualization of the truth label. As shown by Figure 3, the accuracy of the predictions in the "Final Fusion Result" image is far exceed that in the "Raw IPT Output" image.

\begin{figure}[htbp]
\centering
\includegraphics[width=8.1cm]{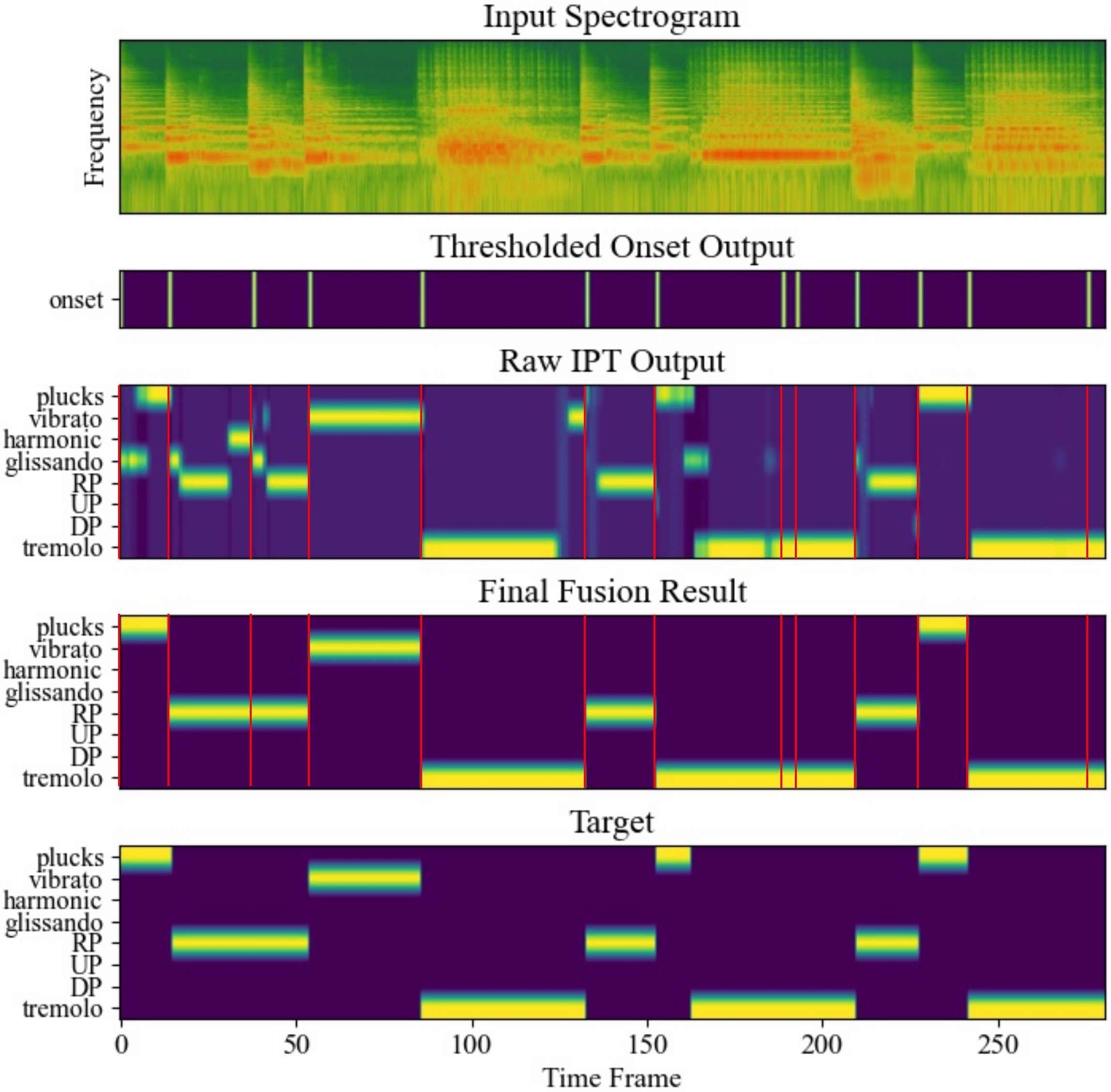}
\caption{Inference on an audio in the test set. The log mel-spectrogram input, the thresholded onset output, the raw IPT output, the final fusion result and the target are shown from top to bottom. The vertical red lines in the third and fourth images are the onset predictions.}
\end{figure}

\subsection{Ablation Studies}

We conducted an ablation study where the performance of \texttt{FCN+Onsets} is compared with  the following five models (namely \texttt{CNN+Onsets}, \texttt{No Onset Fusion}, \texttt{Onset3×3}, \texttt{No Skip Connection}, \texttt{No Weighted Loss}). In \texttt{CNN+Onsets}, we used the same CNN architecture for the IPT detector as presented in the onset detector instead of FCN. The result of \texttt{No Onset Fusion} is the output of the IPT detector without the decision fusion with the onset detector. In \texttt{Onset3×3}, we replaced the whole first convolutional layer of the onset detector by  3×3 filters. In \texttt{No Skip Connection}, we remove the skip connection part of the \texttt{FCN+Onsets} model. \texttt{No Weighted Loss} is the model that uses ordinary binary cross entropy (BCE) Loss in the onset detector. 

The results show the importance of the onset information - \texttt{No Onset Fusion} results in a significant 11.49\% decrease in the frame-level accuracy and a 45.29\% decrease in the note-level F1-score. Our model can locate the IPT boundaries precisely and rectify some mispredicted frames for IPTs through fusing the IPT results with the note onset information. \texttt{CNN+Onsets} results in decreases in all metrics. It shows the FCN can better distinguish different IPTs than CNN. \texttt{No Skip Connection} shows a 5.52\% decrease in frame-level accuracy and a 7.37\% decrease in note-level F1-score. This proves the effectiveness of our skip-connection operation. \texttt{No Weighted Loss} has a little rise in note precision but decrease in all other metrics that shows weighted loss can better balance the precision and recall. 

\section{Conclusion}
In this paper, we create a new dataset consisting of abundant Guzheng playing technique clips from multiple sound banks and real-world recordings. We propose an end-to-end Guzheng playing technique detection system using FCN that can be tested on variable-length audio. We propose a new decision fusion method where an onset detector is trained to divide an audio into several notes and the predicted IPTs are added frame by frame inside each note during fusion. The final output of each note is the IPT class with the highest probability within the note. Our method outperforms existing works by a large margin, which proves that our model is effective and offers good generalization capabilities that transfer well between different training and test sets. Moreover, by visualising the results, we find that the onset information is crucial for the Guzheng playing technique detection task.

In different genres of Guzheng music, there are slight differences in the same playing technique. In the future, we will continue to expand our dataset and add real-world Guzheng music pieces from different genres. Moreover, Guzheng is polyphonic and even one note may have more than one playing techniques. We will investigate how to detect more complicated playing techniques.

\section{ACKNOWLEDGEMENT}
This work was supported by National Key R\&D Program of China (2019YFC1711800), NSFC (62171138). Wei Li, Fan Xia and Yi Yu are corresponding authors of this paper.

% For bibtex users:
\bibliography{ISMIRtemplate}

% For non bibtex users:
%\begin{thebibliography}{citations}
% \bibitem{Author:17}
% E.~Author and B.~Authour, ``The title of the conference paper,'' in {\em Proc.
% of the Int. Society for Music Information Retrieval Conf.}, (Suzhou, China),
% pp.~111--117, 2017.
%
% \bibitem{Someone:10}
% A.~Someone, B.~Someone, and C.~Someone, ``The title of the journal paper,''
%  {\em Journal of New Music Research}, vol.~A, pp.~111--222, September 2010.
%
% \bibitem{Person:20}
% O.~Person, {\em Title of the Book}.
% \newblock Montr\'{e}al, Canada: McGill-Queen's University Press, 2021.
%
% \bibitem{Person:09}
% F.~Person and S.~Person, ``Title of a chapter this book,'' in {\em A Book
% Containing Delightful Chapters} (A.~G. Editor, ed.), pp.~58--102, Tokyo,
% Japan: The Publisher, 2009.
%
%
%\end{thebibliography}
\end{CJK*}
\end{document}